\documentclass[aps,preprint,showpacs,groupedaddress]{revtex4}  
\usepackage{graphicx}  
\usepackage{dcolumn}   
\usepackage{bm}        
\usepackage{amssymb}   

\usepackage{xspace}
\def\d0{D\O}

\def\LQ{$LQ_3$}
\def\mlq{$m_{LQ_3}$}

\def\met{\mbox{$\rlap{\kern0.15em/}E_T$}}
\def\pbarp{$p \overline p$}
\def\etal{{\sl et al.}}
\def\ifb{fb$^{-1}$}
\def\ptvis{$p_T^\tau$}
\def\ettau{$E_T^\tau$}
\def\pttrk{$p_T^{\rm trk}$}

\def\er{$\pm$}
\def\xsth{$\sigma_{\rm th}$}

\begin{document}


\hspace{5.2in} \mbox{Fermilab-Pub-08/201-E}

\title{Search for third generation scalar leptoquarks decaying to $\tau b$}

%
\author{V.M.~Abazov$^{36}$}
\author{B.~Abbott$^{75}$}
\author{M.~Abolins$^{65}$}
\author{B.S.~Acharya$^{29}$}
\author{M.~Adams$^{51}$}
\author{T.~Adams$^{49}$}
\author{E.~Aguilo$^{6}$}
\author{M.~Ahsan$^{59}$}
\author{G.D.~Alexeev$^{36}$}
\author{G.~Alkhazov$^{40}$}
% ...

\affiliation{\vspace{0.1 in}(The D\O\ Collaboration)\vspace{0.1 in}}
% ...
\affiliation{$^{6}$University of Alberta, Edmonton, Alberta, Canada,
                Simon Fraser University, Burnaby, British Columbia, Canada,
                York University, Toronto, Ontario, Canada, and
                McGill University, Montreal, Quebec, Canada}
% ...
\affiliation{$^{29}$Tata Institute of Fundamental Research, Mumbai, India}
% ...
\affiliation{$^{36}$Joint Institute for Nuclear Research, Dubna, Russia}
% ...
\affiliation{$^{40}$Petersburg Nuclear Physics Institute,
                St. Petersburg, Russia}
% ...
\affiliation{$^{49}$Florida State University, Tallahassee, Florida 32306, USA}
% ...
\affiliation{$^{51}$University of Illinois at Chicago,
                Chicago, Illinois 60607, USA}
% ...
\affiliation{$^{59}$Kansas State University, Manhattan, Kansas 66506, USA}
% ...
\affiliation{$^{65}$Michigan State University,
                East Lansing, Michigan 48824, USA}
% ...
\affiliation{$^{75}$University of Oklahoma, Norman, Oklahoma 73019, USA}
% ...

\date{\today}

\begin{abstract}
We have searched for 
third generation
leptoquarks (\LQ ) using 1.05 \ifb ~of 
data collected with the D0 detector at the Fermilab Tevatron 
Collider operating at $\sqrt s = 1.96$ TeV.  We set a 95\% C.L. lower limit of 
210~GeV on the mass of a scalar \LQ ~state decaying 
solely to a $b$ quark and a $\tau$ lepton.
\end{abstract}


\maketitle

The standard model (SM) provides a good description of experimental
data to date, but
fails to address the disparity between the electroweak scale and the much 
higher grand unification or Planck scale.  Models invoking new strong
coupling sectors~\cite{strongcoupling}, 
grand unification~\cite{grandunification}, 
superstrings~\cite{strings}, or 
quark-lepton compositeness~\cite{composite} 
may alleviate this problem.  In these models, new leptoquark
particles ($LQ$) carrying both lepton number and color charge quantum numbers 
may arise.  
The observed suppression of flavor changing neutral currents implies that 
a particular $LQ$ state should couple only to quarks and leptons of the same
fermion generation.  Thus the third generation $LQ$
(\LQ) will decay only into
a $b$ or $t$ quark and a $\tau$ or $\nu_\tau$, depending on the \LQ ~electric 
charge.
At the Fermilab Tevatron \pbarp ~Collider, leptoquarks can be 
pair-produced through gluon-gluon fusion
and $q \overline q$ annihilation 
with standard QCD color interactions.
The charge 4/3 \LQ ~decays to $\tau^+ \overline b$
with a branching ratio (BR) of 1,
whereas the charge 2/3 \LQ ~ decays to $\tau^+  b$ with coupling constant 
$\beta$ and to $\overline \nu_\tau  t$ with coupling 
$(1-\beta)$ (and charge conjugates for $\overline{LQ_3}$ decays). 
For the $\overline \nu_\tau  t$ decay, the BR $= (1-\beta)\times f_{\rm PS}$ 
is further suppressed
by the phase space factor $f_{\rm PS}$ due to the large top quark mass. 
  
In Run II, the D0 collaboration set 
a lower mass limit of 229~GeV~\cite{dzruntwo} for the charge 1/3 
\LQ ~$\rightarrow \nu_\tau \overline b$. 
Here we present new limits on the mass of leptoquarks with charge 
4/3 with decays \LQ $\rightarrow \tau \overline b$ and 
charge 2/3 with
decays \LQ $\rightarrow \tau b$. The best previous limit for this
channel is 99~GeV~\cite{cdfrunoneb,opal}. 
For pair production, both \LQ ~charge states lead to 
the final state $\tau^+\tau^- b\overline b$.  We identify one of the $\tau$
leptons through its decay $\tau \rightarrow \mu \nu_\mu \nu_\tau$ and the
other through its hadronic decays. The presence of jets from 
$b$ quarks is signalled by 
tracks displaced from the primary vertex. The final state
sought is thus two $b$ jets, $\mu$, $\tau$, and missing transverse
energy (\met ).   

The D0 detector~\cite{d0detrun1,d0detrun2} has a central
tracking volume with a silicon microstrip vertex detector 
(pseudorapidity coverage $|\eta|<3$)
and a scintillating fiber tracker ($|\eta|<2.5$) within
a 2 T solenoidal magnet; a uranium/liquid-argon calorimeter 
($|\eta|< 4.2$); and a surrounding muon 
identification system ($|\eta|<2$),
with tracking chambers and scintillators before and after solid iron toroid
magnets. Events are
selected using a suite of triggers requiring either a single muon or a muon
in association with jets. 
This analysis is performed using 1.05 \ifb~of data collected in Run II. 

Muon candidates are required to have hits in the muon system  matched to 
a track candidate with $p_T > 15$~GeV 
and $|\eta |<2$, and are required to
extrapolate to within 1.5 cm of the reconstructed primary 
vertex along the beam axis.  Cosmic ray muons
are removed using the muon scintillation counter timing.   Muon 
candidates are required to be isolated from nearby particles by requiring
a calorimeter energy deposit of less than 2.5~GeV 
within a hollow cone of $0.1 < {\cal R}
< 0.4$ centered on the muon direction, and less than 2.5~GeV associated 
with tracks (excepting the muon track) within $\cal R$ $< 0.5$.  
Here, $\cal R$ $=\sqrt{(\Delta \phi)^2+(\Delta \eta)^2}$ 
is the distance in $\eta$-$\phi$ space between objects.

We identify three types of tau candidate motivated by the decays: 
(1) $\tau^\pm \rightarrow \pi^\pm \nu$, 
(2) $\tau^\pm \rightarrow \pi^\pm \pi^0 {\rm 's} ~\nu$, and
(3) $\tau^\pm \rightarrow \pi^\pm \pi^\pm \pi^\mp (\pi^0 {\rm 's)} \nu$.
The corresponding selections~\cite{taunn} 
for the three types are based on  
tracks with transverse momenta \pttrk ~$>1.5$~GeV and energy 
clusters in the electromagnetic (EM) calorimeter, both 
within a cone of ${\cal R}$ $<0.5$.  The visible transverse momentum of
a tau candidate, \ptvis , is constructed from the calorimeter transverse
energy (\ettau), corrected by track information where warranted. 
 The tau selections are  
(1) a single isolated
track with transverse momentum \pttrk ~$>15$~GeV
and no nearby electromagnetic energy cluster; 
(2) a single isolated track with \pttrk $>7$~GeV with an 
associated EM cluster; and 
(3) two or more tracks, with at least one having \pttrk ~$>7$~GeV, with or
without associated EM clusters.  
Tau candidates must have 
\ptvis ~above
15~GeV for types 1 and 2, and above 20~GeV for type 3. 
For type~1
candidates, we require \pttrk/\ettau ~$\ge 0.7$
to reduce contributions from $\tau^\pm \rightarrow \pi^\pm \pi^0$'s in 
calorimeter regions with poor EM particle identification
and \pttrk/\ettau ~$\le 2.0$ to reduce backgrounds from muons.
A neural network~\cite{taunn} 
is formed for each $\tau$ type using input variables
such as isolation and the transverse and longitudinal shower profiles
of the calorimeter energy depositions associated with
the tau candidate.  The networks give an output 
variable ${\cal N}_i$ 
for $\tau$-type $i$.  We require $\cal N$$_1$, $\cal N$$_2$ and $\cal N$$_3$ 
to exceed 0.9, 0.9 and 0.95, corresponding to about 70\% efficiency with 
$\ge 90$\% rejection of fake jets.

\begin{figure*}[t]
\begin{center}
\includegraphics[width=0.32\textwidth]{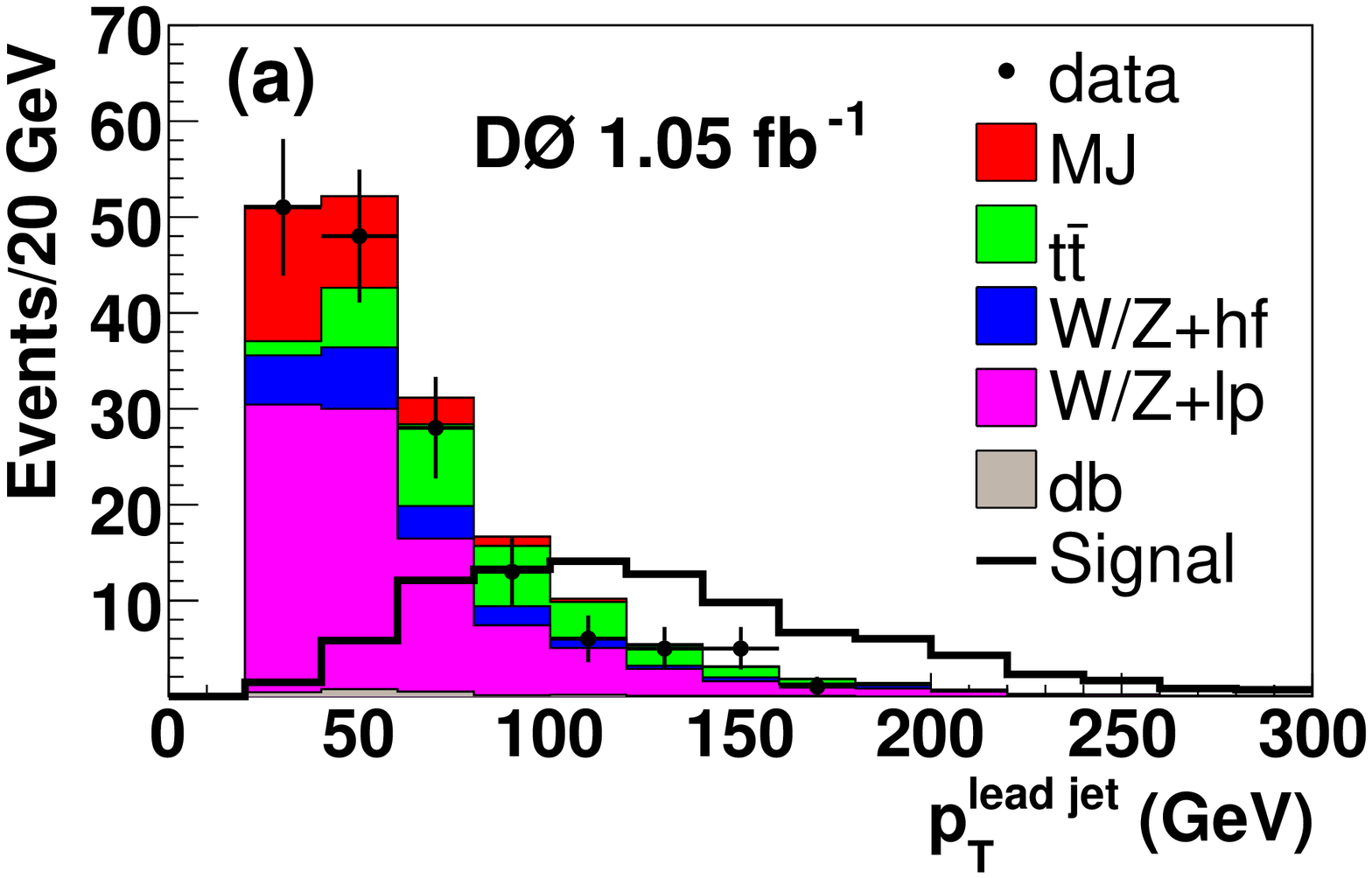}
\includegraphics[width=0.32\textwidth]{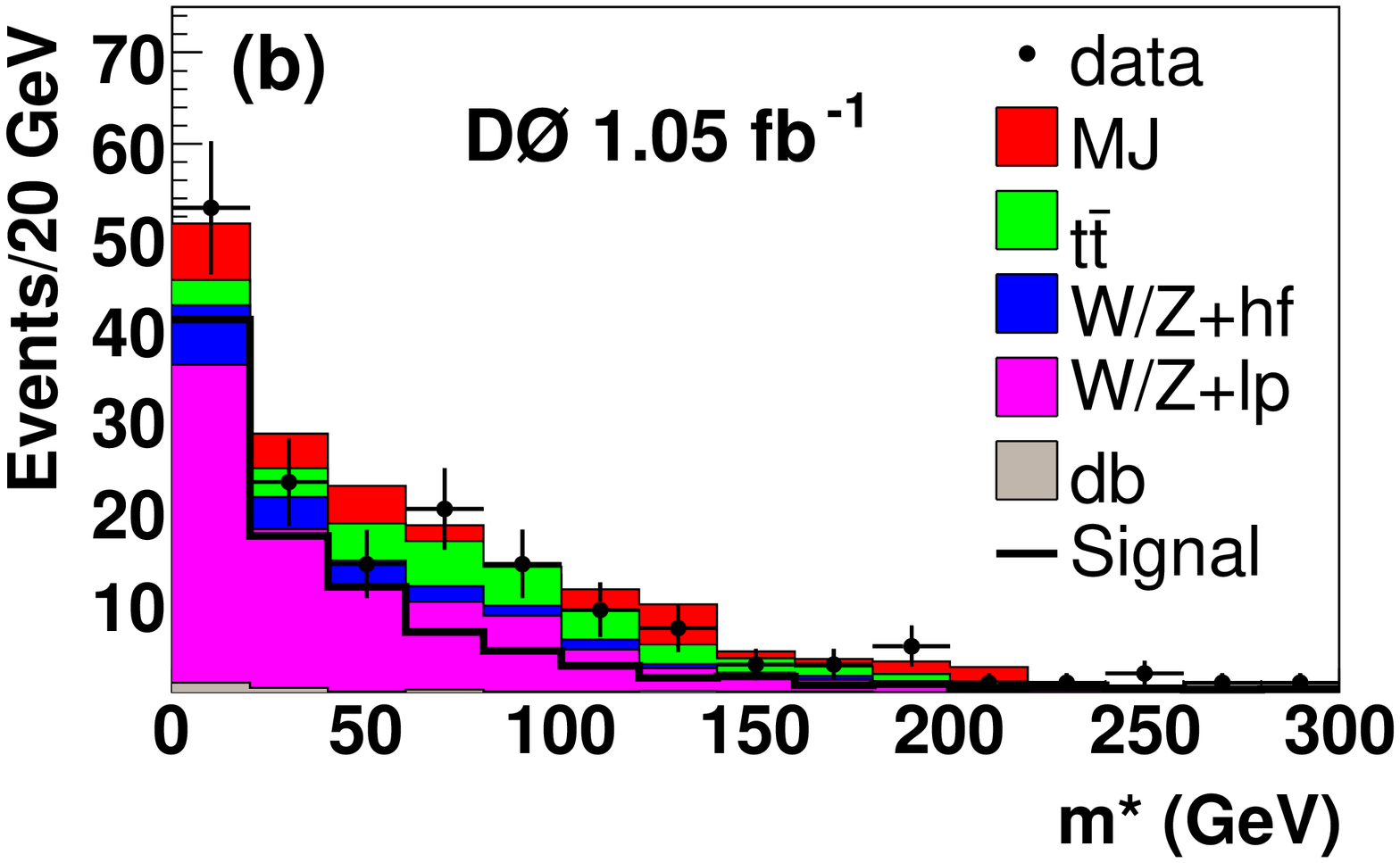}
\includegraphics[width=0.32\textwidth]{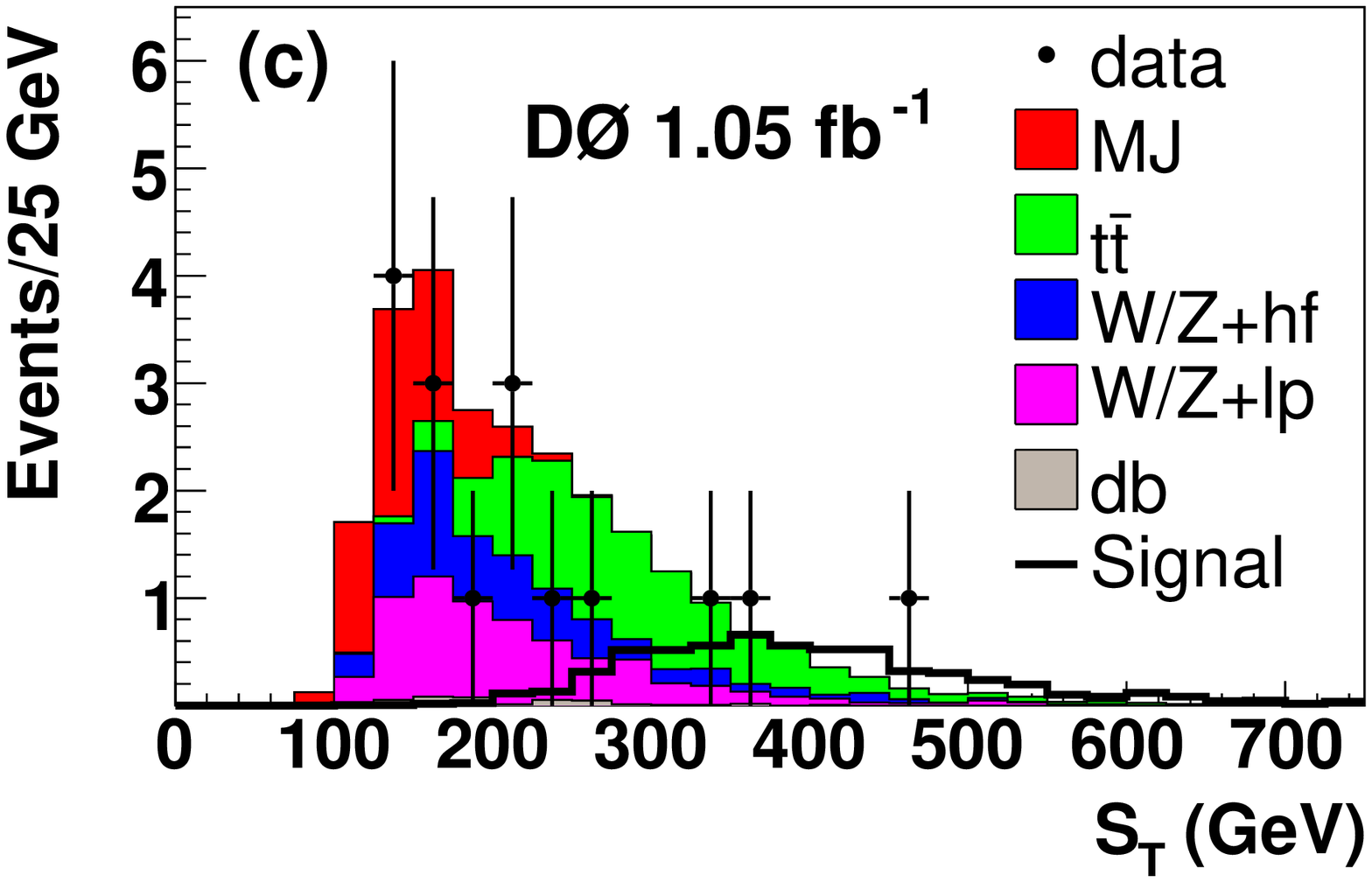}
\caption{\label{fig:fig1} Comparisons of data and the sum of 
backgrounds
for (a) $p_T$ of the highest $p_T$ jet after preselection, 
(b) $m^*$ after preselection, and (c) $S_T$ for 
the 1 and $\ge 2$ b tagged samples combined. We denote 
the diboson contribution as ``db'', heavy quarks ($b$, $c$)
as ``hf'' and light partons ($u$, $d$, $s$ and gluons) as ``lp''.
The \LQ ~signal is shown for \mlq = 200~GeV, multiplied by 10 in (a) and (b) 
and without scaling in (c).  (color online)} 
\end{center}
\end{figure*}

We reconstruct jets using calorimeter energy deposits within a cone radius
of 0.5~\cite{jetalgorithm} and correct to the particle level using
a jet energy scale 
correction (JES).  Jets containing a muon  are 
further corrected for the muon and average neutrino energies. Jets are required
to have $p_T > 20$~GeV ($>25$~GeV for the highest $p_T$ jet) and 
$|\eta|<2.5$ relative to the center of the detector.  
We calculate \met ~from the transverse plane
vector sum of calorimeter energy deposits, corrected for observed muons
and for the jet and $\tau$ energy scale corrections.

We tag jets as $b$-jet candidates using a neural network 
algorithm~\cite{btagger} employing track
impact parameters, significance of track displacement from the 
primary vertex, vertex
mass, and number of tracks associated with a secondary vertex.   The 
selection on the
neural network output is optimized for the best \LQ ~sensitivity and has
72\% efficiency for $b$ jets, with a misidentification probability for
light quark jets of 6\%.  

Events are preselected with the requirements that there is only 
one isolated muon, and 
at least two jets with
$\cal R$ $>0.5$ relative to the 
$\mu$ or $\tau$ candidates. If more than one $\tau$ 
candidate is found, the one with the largest \ptvis ~is chosen.
We require no electrons with $p_T>12$~GeV.

The \LQ ~signal is simulated for $m_{\rm LQ_3} = 120$ to 220~GeV 
in 20~GeV steps, 
using the {\footnotesize PYTHIA}~\cite{pythia} Monte Carlo (MC) 
generator and CTEQ6L1~\cite{cteq} parton distribution functions
(PDF). The normalization at 
next-to-leading order (NLO) is taken from~\cite{kramer}.  
The $t\overline t$ and  $W/Z$ boson+jets 
backgrounds are simulated with the {\footnotesize ALPGEN} 
MC generator~\cite{alpgen}, 
with {\footnotesize PYTHIA} used for parton showering and fragmentation. 
The $t \overline t$ cross section is normalized to 
the NNLO cross section~\cite{ttbarxs} with top quark mass $m_t=175$~GeV 
and the $W/Z$+jets cross sections are
normalized to the $W/Z$ inclusive NLO cross section~\cite{mcfm}. 
The $WW$, $WZ$, and $ZZ$ diboson backgrounds are generated using 
{\footnotesize PYTHIA} and normalized to the NLO cross 
sections~\cite{mcfm}.  The $\tau$ 
polarization and decays for all processes are simulated with 
{\footnotesize TAUOLA}~\cite{tauola}. The simulated 
events are processed through a {\footnotesize GEANT}~\cite{geant} 
detector simulation and the standard D0 event reconstruction. They  
are further corrected for differences between data and MC simulation 
in the identification efficiencies for muons, electrons and jets, $Z$ boson
$p_T$, the distribution of primary vertices along the beam axis, 
jet energy scale and resolution, $b$-jet tagging, and the effect
of additional minimum bias interactions.  
The trigger efficiency applied to the simulated events is 
measured as a function of muon and jet azimuthal angle $\phi$ and 
$\eta$, and is appropriately averaged using the instantaneous luminosity
in each data collection epoch.  

We determine the multijet (MJ) background from two data samples, after 
subtracting the simulated SM backgrounds for both.  The 
signal (SG) sample is that obtained from the preselected data 
discussed above.
The enhanced  background (BG) sample uses the preselection 
cuts, except that the muon track 
and calorimeter isolation requirements are reversed, and the $\tau$ 
identification requires $\cal N$$_i <0.8$.  The shapes of the BG kinematic
distributions agree well with those
for the SG sample and provide the shape of the MJ background. 
We subdivide both SG and BG samples into opposite
sign (OS) and same sign (SS) subsets according to whether the observed $\mu$ 
and $\tau$ charges are opposite or the same, with numbers of events, 
$N_C^Q$ ($C$=SG,BG) ($Q$=OS,SS).  The MJ background is computed as 
$N_{\rm SG}^{\rm OS} = f \times N_{\rm SG}^{\rm SS}$, 
where the MJ normalization factor is 
$f = N_{\rm BG}^{\rm OS}/N_{\rm BG}^{\rm SS}$.  
The factor $f$ is observed to be close to 1 and independent of $p_T^\mu$ and
$p_T^\tau$.  There is negligible \LQ ~signal in the SS BG subsample.
 
Further analysis uses the OS preselected events.
Figure 1(a) shows the $p_T$ distributions of the data and the 
sum of all backgrounds
for the highest $p_T$ jet.  The agreement for
this and other kinematic distributions is good.  Figure 1(b) shows the data 
and background distributions for a variable related to the $W$ boson mass, 
defined as $m^*= \sqrt{2E^\mu E^\nu (1-\cos\Delta \phi)}$ where the estimated
neutrino energy is
$E^\nu =$ \met $(E^\mu/p_T^\mu)$, and $\Delta \phi$ is the azimuthal 
angle between the muon and \met ~directions. The $t\overline t$ and 
$W$ + jets backgrounds 
contain a real $W$ boson and have a high value of $m^*$, whereas the \LQ 
~signal tends to have small $m^*$.  Based on the expected \LQ ~mass limit
from MC studies, we require $m^* < 60$~GeV.   

The jets in the event sample after the $m^*$ cut are subjected to the 
$b$-tagging algorithm
and subsets are formed with exactly one tagged $b$ jet and with 
$\ge 2~b$ jet tags.  The numbers of events in the OS preselection sample, 
after the $m^*$ requirement, and the 
1 and $\ge 2$ $b$-tagged jet subsamples, are shown in 
Table~\ref{tab:yields}.  

We define the variable $S_T$ as the scalar 
sum of the transverse momenta of $\mu$, $\tau$, the two highest $p_T$ jets,
and  \met .  The
\LQ ~signal is expected to have higher values of $S_T$ than the
 background processes.  Figure 1(c) shows the distributions of $S_T$ for 
data and expected background, for the 1 $b$-tagged jet and
$\ge 2~b$-tagged jet samples combined.  We observe no excess 
above the expected backgrounds.

\begin{table}[bt]
\caption{\label{tab:yields}
Number of events for data and estimated backgrounds at the preselection level,
after the $m^*$ cut (before $b$-jet tagging), and for the 
 $1~b$-tag and $\ge 2~b$-tag subsets. 
Light partons ($u,d,s, g$) are denoted as ``lp''.   Also shown is the expected 
number of signal events for \mlq ~= 200~GeV. The uncertainties 
shown are statistical.}
\begin{tabular}{lrrrr} 
\hline
\hline
 Source & Preselection & $m^* < 60$ GeV &1 $b$-tag & $\ge$ 2 $b$-tag \\ \hline
$W+\rm{lp}$          & 29.8\er 1.8 & 11.2\er 1.0 & 1.0\er 0.4 &  $<0.1$  \\
$W+c\overline c$ &  4.0\er 0.4 & 1.5\er 0.2 & 0.4\er 0.1  &  $<0.1$  \\
$W+b\overline b$ &  2.2\er 0.2 & 0.8\er 0.1 & 0.4\er 0.1  &  $<0.1$  \\
$Z+{\rm lp}$          & 64.0\er 0.7 & 55.3\er 0.7 & 5.0\er 0.2 & 0.1\er 0.0 \\
$Z+c\overline c$ &  8.3\er 0.5 & 7.3\er 0.5  & 1.7\er 0.2 & 0.1\er 0.1 \\
$Z+b\overline b$ &  4.4\er 0.2 & 3.8\er 0.1 & 1.8\er 0.1  & 0.4\er 0.1 \\
$t\overline t$  & 29.8\er 0.3 &  10.6\er 0.1 &  5.2\er 0.1  & 3.1\er 0.1 \\
Diboson         & 2.0\er 0.2  & 1.5\er 0.1  & 0.3\er 0.1  &  $<0.1$  \\
MJ              & 25.2\er 7.6 & 17.2\er 5.6 & 4.0\er 2.5  & 0.8\er 1.0   \\ 
Sum Bknd & 169.6\er 7.9 & 109.2\er 5.7 & 19.6\er 2.5 & 4.8\er 1.0\\ \hline
Data            &   157  &   94   &   15   &   1      \\ \hline
LQ pair signal & 9.0\er 0.2 & 7.4\er 0.1 & 3.4\er 0.1 & 2.6\er 0.1 \\
\hline \hline
\end{tabular}
\end{table}

The systematic uncertainty for the  
luminosity determination (6.1\%) is taken from~\cite{lumi}. 
Calibration data sets are used
to determine the uncertainties on the trigger efficiency (3\%)
and  on the reconstruction, 
identification and isolation efficiencies for the $\mu$, $\tau$, and
 jets (7\%).
The MC acceptance uncertainties due to the jet energy  
uncertainty are found to be 6 -- 9\% by varying the JES by 
$\pm $ one standard deviation~\cite{incljet}.   
The uncertainties on the tagging rates for heavy flavor and light parton
jets result in systematic uncertainties on the signal acceptance  
(7.5\%), and on the $W/Z$ + heavy flavor jets background (7.5\%) and $W/Z$ + 
light parton jets background (15\%)~\cite{btagger}.  
The MJ background uncertainty (15\%) is determined by using  
independent MJ data samples in which
either the $\mu$ isolation cuts or the $\tau$ neural network cuts 
(but not both) are reversed.  
The $t\overline t$ cross section uncertainty (18\%)
incorporates the estimated 
theoretical dependence on the renormalization and factorization 
scales~\cite{ttbarxs}, the uncertainty 
on $m_t$ and the uncertainty
due to the PDF choice.
The diboson production cross section uncertainty (6\%) and the 
$W/Z$ + jets cross section uncertainties (22\%) are
estimated using {\footnotesize MCFM}~\cite{mcfm}.  
 
We compute the 95\% C.L. upper limits on the signal cross section
as a function of \mlq ~
using the modified frequentist method~\cite{junk} as implemented 
in~\cite{fisher}.  Negative log-likelihood ratio (LLR) test statistics 
are formed and combined
from the $S_T$ distributions for 1 and $\ge 2~b$-tagged samples in 
simulated pseudo-experiments, under 
the background only (LLR$_{\rm b}$) and signal plus 
background (LLR$_{\rm s+b}$) hypotheses.  
We integrate LLR$_{\rm b}$  (LLR$_{\rm s+b}$) above the LLR value observed in
data to obtain confidence levels CL$_{\rm b}$  (CL$_{\rm s+b}$).  The 
\LQ ~cross section is varied until the ratio CL$_{\rm s} =$  
CL$_{\rm s+b}$/CL$_{\rm b}$ equals 0.05.
The resulting expected and observed  limits are shown in Fig. 2,
together with the theoretical cross section (\xsth ) assuming  
BR = 1. The observed cross section
limit is within one standard deviation of the expected limit
for \mlq ~$\approx 200$~GeV, and within two standard deviations for
all masses. The uncertainty on \xsth ~is obtained by varying 
renormalization and factorization scales
by a factor of two above and below the central value of \mlq ~and
by taking into account the uncertainties in the PDFs~\cite{cteq,sigtheory}. 
The intersection of the observed cross section limit and the central \xsth 
~as a function of \mlq 
~yields the exclusion of \mlq ~$> 210$~GeV (for $\beta=1$), and 
at the one standard deviation lower value of \xsth ~we find  
\mlq ~$> 201$~GeV, both at the 95\% C.L.  

The dashed line in Fig. 2 indicates
the decrease in the cross section $\times$ BR$^2$ for 
the charge 2/3 \LQ ~$\rightarrow \tau b$ 
when the decay \LQ ~$\rightarrow \nu_\tau t$ becomes
kinematically possible, after including $f_{\rm PS}$  
(for $\beta = 0.5$ and 
$m_t$ = 175~GeV).  In this case,
we obtain \mlq ~$> 207$~GeV for the central \xsth ~ and 
\mlq ~$>201$~GeV for the one standard deviation lower limit of \xsth ,
at 95\% C.L.

\begin{figure}
\begin{center}
\includegraphics[width=0.5\textwidth]{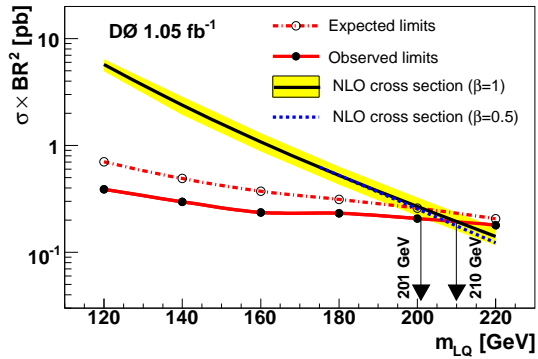}
\caption{\label{fig:fig2} Observed and expected cross section 
limits at the 95\% C.L. of the pair 
production of third generation leptoquarks as 
a function of \mlq . The uncertainty on the theoretical prediction is 
shown with shaded error bands. The theoretical cross section times branching
ratio when $\beta = 1/2$ is shown as the dashed line. (color online)}
\end{center}
\end{figure}

In summary, we have searched for third generation leptoquark pair
production with
decays \LQ ~$\rightarrow \tau b$, and exclude 
\mlq ~$< 210$~GeV at the 95\% C.L., 
assuming the branching fraction for this mode to be one.  This is the most
stringent limit on third generation leptoquarks in this decay channel to date.

%
We thank the staffs at Fermilab and collaborating institutions, 
and acknowledge support from the 
DOE and NSF (USA);
CEA and CNRS/IN2P3 (France);
FASI, Rosatom and RFBR (Russia);
CNPq, FAPERJ, FAPESP and FUNDUNESP (Brazil);
DAE and DST (India);
Colciencias (Colombia);
CONACyT (Mexico);
KRF and KOSEF (Korea);
CONICET and UBACyT (Argentina);
FOM (The Netherlands);
STFC (United Kingdom);
MSMT and GACR (Czech Republic);
CRC Program, CFI, NSERC and WestGrid Project (Canada);
BMBF and DFG (Germany);
SFI (Ireland);
The Swedish Research Council (Sweden);
CAS and CNSF (China);
and the
Alexander von Humboldt Foundation (Germany).

\end{document}